\documentclass[prd,superscriptaddress,twocolumn,nofootinbib,showpacs,preprintnumbers,floatfix]{revtex4}

\usepackage{mathrsfs}
\usepackage{amsfonts,amssymb,amsmath}
\usepackage{graphicx,epsfig}
\usepackage{multirow}
\usepackage{verbatim}

\def\fsu5{$\cal{F}$-$SU(5)$}
\def\bfsu5{$\boldsymbol{\mathcal{F}}$-$\boldsymbol{SU(5)}$}

\def\m1half{$M_{1/2}$}
\def\m3half{$M_{3/2}$}
\def\m32{$M_{32}$}

\def\fb{${\rm fb}^{-1}$~}

\def\mt2{$M_{T2}$}
\def\x2{$\chi^2$}
\def\2b{$M_{T2}b$}
\def\sb{$S/\sqrt{B+1}$~}

\begin{document}

\title{Non-trivial Supersymmetry Correlations between ATLAS and CMS Observations}

\author{Tianjun Li}

\affiliation{State Key Laboratory of Theoretical Physics and Kavli Institute for Theoretical Physics China (KITPC),
Institute of Theoretical Physics, Chinese Academy of Sciences, Beijing 100190, P. R. China}

\affiliation{George P. and Cynthia W. Mitchell Institute for Fundamental Physics and Astronomy,
Texas A$\&$M University, College Station, TX 77843, USA}

\author{James A. Maxin}

\affiliation{George P. and Cynthia W. Mitchell Institute for Fundamental Physics and Astronomy,
Texas A$\&$M University, College Station, TX 77843, USA}

\author{Dimitri V. Nanopoulos}

\affiliation{George P. and Cynthia W. Mitchell Institute for Fundamental Physics and Astronomy,
Texas A$\&$M University, College Station, TX 77843, USA}

\affiliation{Astroparticle Physics Group, Houston Advanced Research Center (HARC),
Mitchell Campus, Woodlands, TX 77381, USA}

\affiliation{Academy of Athens, Division of Natural Sciences,
28 Panepistimiou Avenue, Athens 10679, Greece}

\author{Joel W. Walker}

\affiliation{Department of Physics, Sam Houston State University,
Huntsville, TX 77341, USA}

%%%%%%%%%%%%%%%%%%%%%%%%%%%%%%%%%%%%%%%%%%%%%%%%%%%%%%%%%%%%%%%%%%%%%%%%%%%%

\begin{abstract}
We present definite correlations between the CMS 5 \fb all-hadronic search employing the stransverse mass variable $M_{T2}$ and the ATLAS 5 \fb all-hadronic and multijet supersymmetry
(SUSY) searches, suggesting the possibility that both the ATLAS and CMS experiments are already registering a faint but legitimate SUSY signal at the LHC.
We isolate this prospective mutual productivity beyond the Standard Model in the framework of the supersymmetric No-Scale Flipped $SU$(5) grand unified theory,
supplemented with extra vector-like matter (flippons). Evident overproduction is observed in three CMS \mt2 and four ATLAS hadronic and multijet signal regions,
where a \x2 fitting procedure of the CMS 5 \fb \mt2 search establishes a best fit SUSY mass in sharp agreement with corresponding ATLAS searches of equivalently
heightened signal significance. We believe this correlated behavior across two distinct experiments at precisely the same SUSY mass scale
to be highly non-trivial, and potentially indicative of an existing 5 \fb LHC reach into a pervasive physical supersymmetry framework.
\end{abstract}

%%%%%%%%%%%%%%%%%%%%%%%%%%%%%%%%%%%%%%%%%%%%%%%%%%%%%%%%%%%%%%%%%%%%%%%%%%%%


\preprint{ACT-08-12, MIFPA-12-20}

\maketitle

%%%%%%%%%%%%%%%%%%%%%%%%%%%%%%%%%%%%%%%%%%%%%%%%%%%%%%%%%%%%%%%%%%%%%%%%%%%%

The conclusion of the 2011 $\sqrt{s}$ = 7 TeV run at the LHC has yielded several recent analyses by the ATLAS and CMS Collaborations of
the nominal 5 \fb data harvest.  A rigorous cross-examination of these studies within the high-energy framework of
a model known as \fsu5 (See Refs.~\cite{Maxin:2011hy,Li:2011rp,Li:2011fu,Li:2011av,Li:2011ab,Li:2012hm,Li:2011xu} and all references therein),
which combines the No-Scale Flipped SU(5) grand unified theory (GUT) with extra vector-like particles (flippons),
prompted our ensuing suggestion~\cite{Li:2012tr} that early indications of supersymmetry (SUSY) production may have
already been accumulated at the LHC, and isolated, in particular, by ATLAS.
The centrally intriguing aspects of the Ref.~\cite{Li:2012tr} analysis are the demonstration of distinct
correlations in i) the SUSY scale favored by those search techniques displaying noticeable production beyond the data-driven Standard Model (SM)
backgrounds, and ii) a precise upscaling from the event excesses realized within the historical 1 \fb data sets in the
transition to the much larger 4.7 \fb data compilations.
One vital ingredient absent from this analysis was any dynamic contribution by published CMS 5 \fb studies to the subset of
overproductive searches. The task of the present effort is to supplement the previously analyzed ATLAS searches
exhibiting positive signal excesses with results from an intervening study released by CMS that shows similar excess
event activity beyond the SM expectations.

The recent completion of a new 5 \fb analysis by CMS~\cite{SUS-12-002} searching for SUSY in hadronic final states using the stransverse mass variable \mt2
triggered a detailed inspection by our group to discern whether any of these new individual searches demonstrate interesting event overproduction, and whether
any such excesses could be explained within the No-Scale \fsu5 model in a manner consistent with our existing best fit against the ATLAS data~\cite{Li:2012tr}. 
This CMS search tactic segregates the results into two categories, one referred to as \mt2 ($\ge$ 3j), and another dubbed \2b ($\ge$ 4j and $\ge$ 1 b-jet), with two signal regions comprised of low $H_T$ ($750 \le H_T \le 950$ GeV) and high $H_T$ ($H_T > 950$ GeV) statistics, providing four discrete signal regions. Findings reveal that indeed three of the four search regions do show signs of excessive activity, in both the \mt2 and \2b low $H_T$ signal regions with a cut on the variable \mt2 implemented at 200 GeV, and also in the \mt2 high $H_T$ signal region with a cut on the variable \mt2 of 150 GeV. An application of the cut at 200 GeV on the variable \mt2 in the low $H_T$ signal region may be motivated to ensure suppression of all QCD background. As a quantitative measure of comparison of diverse LHC search methodologies, we utilize the signal significance metric \sb to assess the strength of the event production exceeding the data-driven background estimates.
For the CMS \mt2 and \2b searches noted above, we find \sb = 2.22 for the \mt2 low $H_T$, \sb = 2.06 for the \2b low $H_T$, and \sb = 1.36 for the \mt2 high $H_T$. 

In order to broaden the spectrum of overproductive 5 \fb searches at the LHC which we include in our analysis, we opt here for a softer lower boundary on signal
significance of \sb = 1.0, such that a more evenly distributed number of admissible active searches from both ATLAS and CMS can be evaluated and potentially linked. As a consequence, in addition to the three CMS \mt2 and \2b searches noted above, we further find four ATLAS searches that surpass our minimum threshold for signal significance: the 7-jet pT$>$80 GeV (7j80) (\sb = 2.07) and 8-jet pT$>$55 GeV (8j55) (\sb = 1.18) cases of the ATLAS multijets~\cite{ATLAS-CONF-2012-037}, and the SRC Tight (\sb = 3.22) and SRE Loose (\sb = 2.65) cases of the ATLAS hadronic 0-lepton search~\cite{ATLAS-CONF-2012-033}. We believe that the resulting four 5 \fb ATLAS and three 5 \fb CMS search strategies represent a reasonable comparative test bed 
from which to probe whether notable correlations are emerging between the distinct LHC experiments.

\begin{figure*}[htp]
        \centering
        \includegraphics[width=0.8\textwidth]{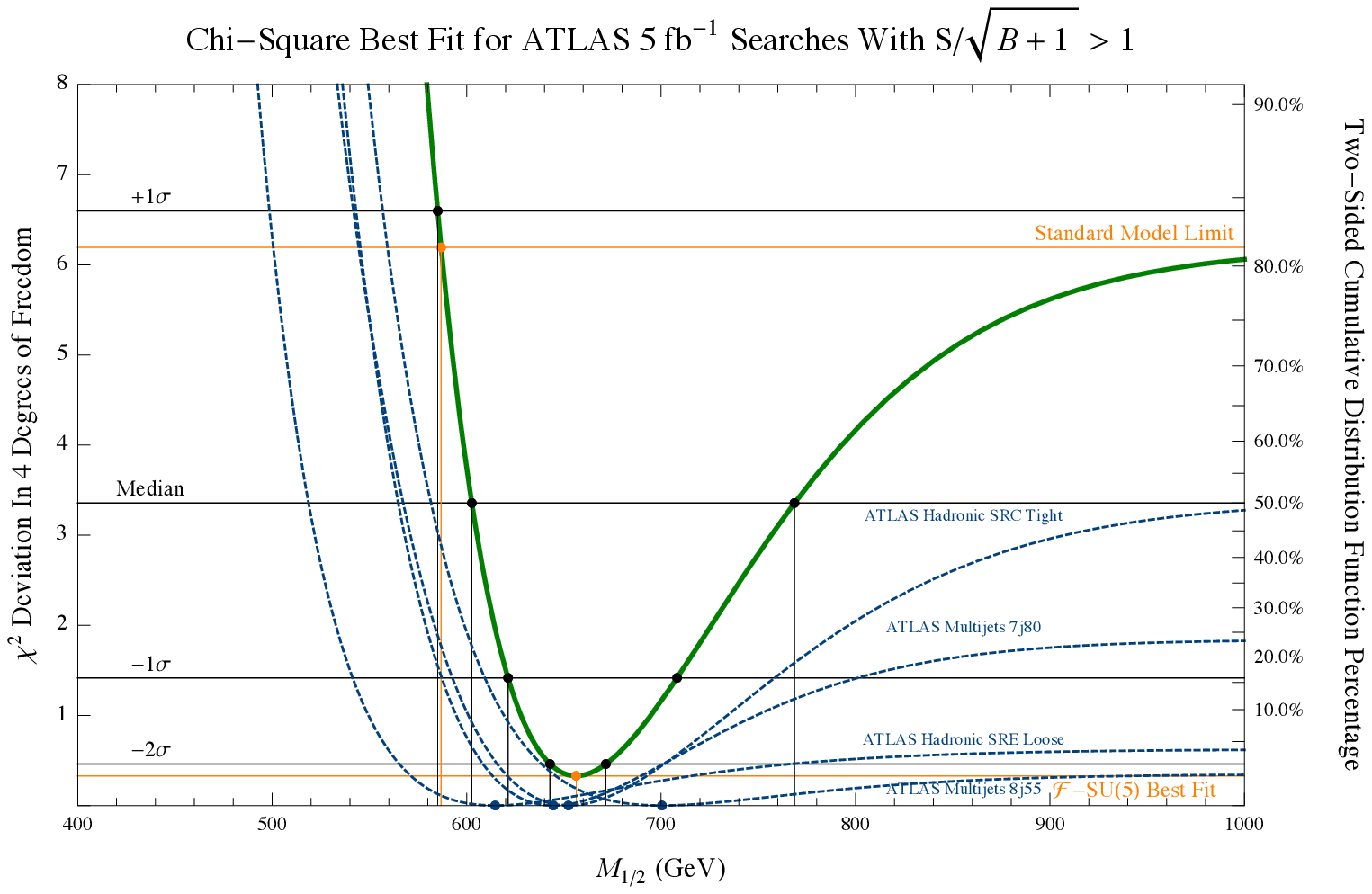}
        \includegraphics[width=0.8\textwidth]{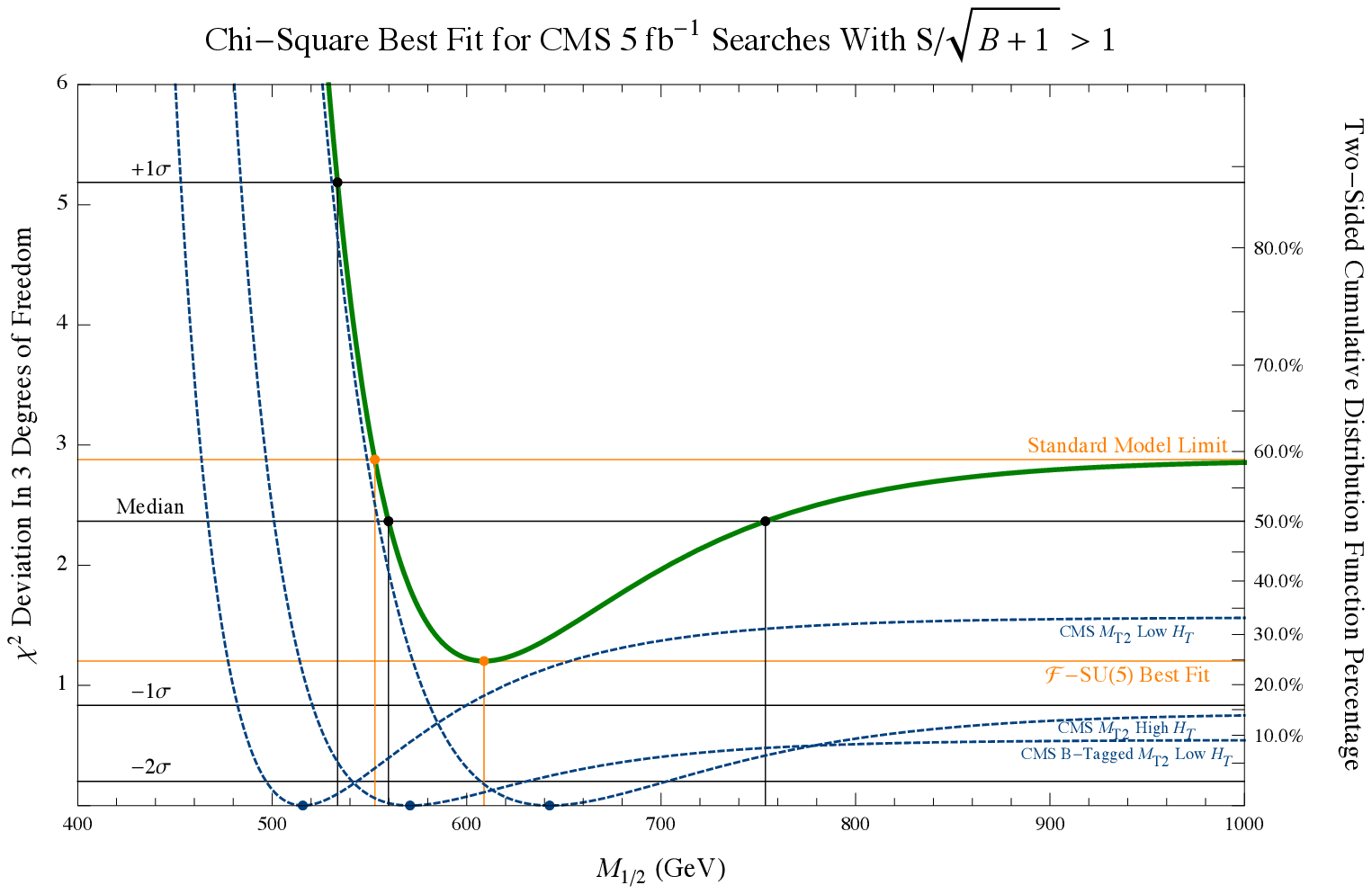}
        \caption{We depict the $\chi^2$ analysis of the ATLAS 4.7 \fb 7j80, 8j55, SRC Tight, and SRE Loose Multijet search strategies from Refs.~\cite{ATLAS-CONF-2012-037,ATLAS-CONF-2012-033} in the upper pane, and the CMS 4.73 \fb \mt2 low $H_T$, \2b low $H_T$, and \mt2 high $H_T$ of Ref.~\cite{SUS-12-002} in the lower pane. The thin dotted blue lines correspond to the individual $\chi^2$ curves for each event selection, which are summed into the thick green cumulative multi-axis $\chi^2$ curves.  These searches are selected for the exhibition of a signal significance $S/\sqrt{B+1}$ greater than 1 for the 5 \fb class studies.  A direct visual inspection of the correspondence of the signal
strength and the fluctuation of the $\chi^2$ minimum with increased luminosity is thus facilitated. Remarkably, we
observe extreme stability in the favored mass scale between the two independent LHC experiments. Such correlations across diverse experiments continue to indicate the presence of a non-random structure. It remains highly improbable that the source of the SUSY mass correlation is arbitrary fluctuations of the data-driven background simultaneously inflicting the same point in the targeted SUSY spectrum.}
        \label{fig:CSBF}
\end{figure*}

\begin{figure*}[htb]
        \centering
        \includegraphics[width=0.8\textwidth]{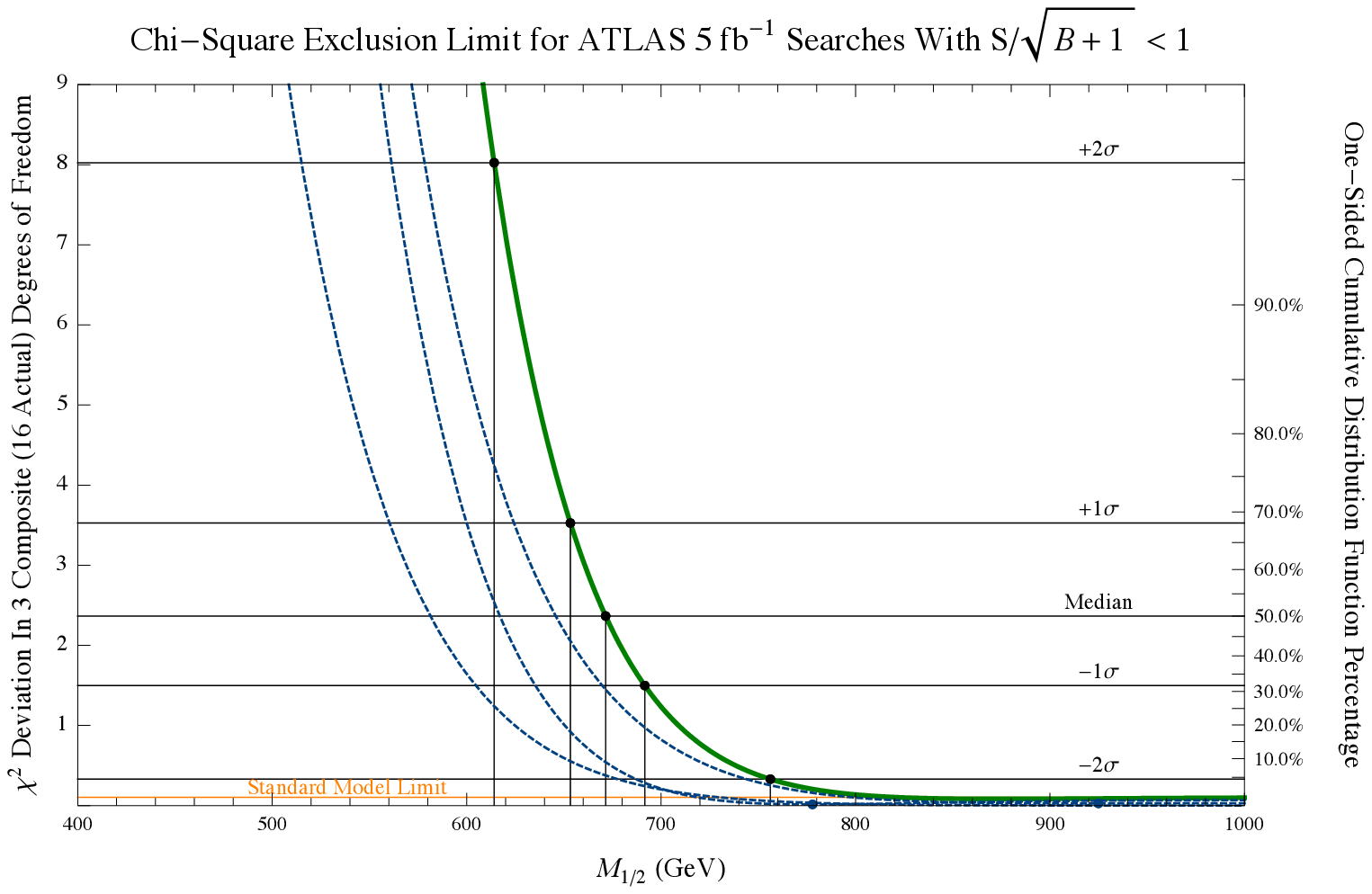}
        \includegraphics[width=0.8\textwidth]{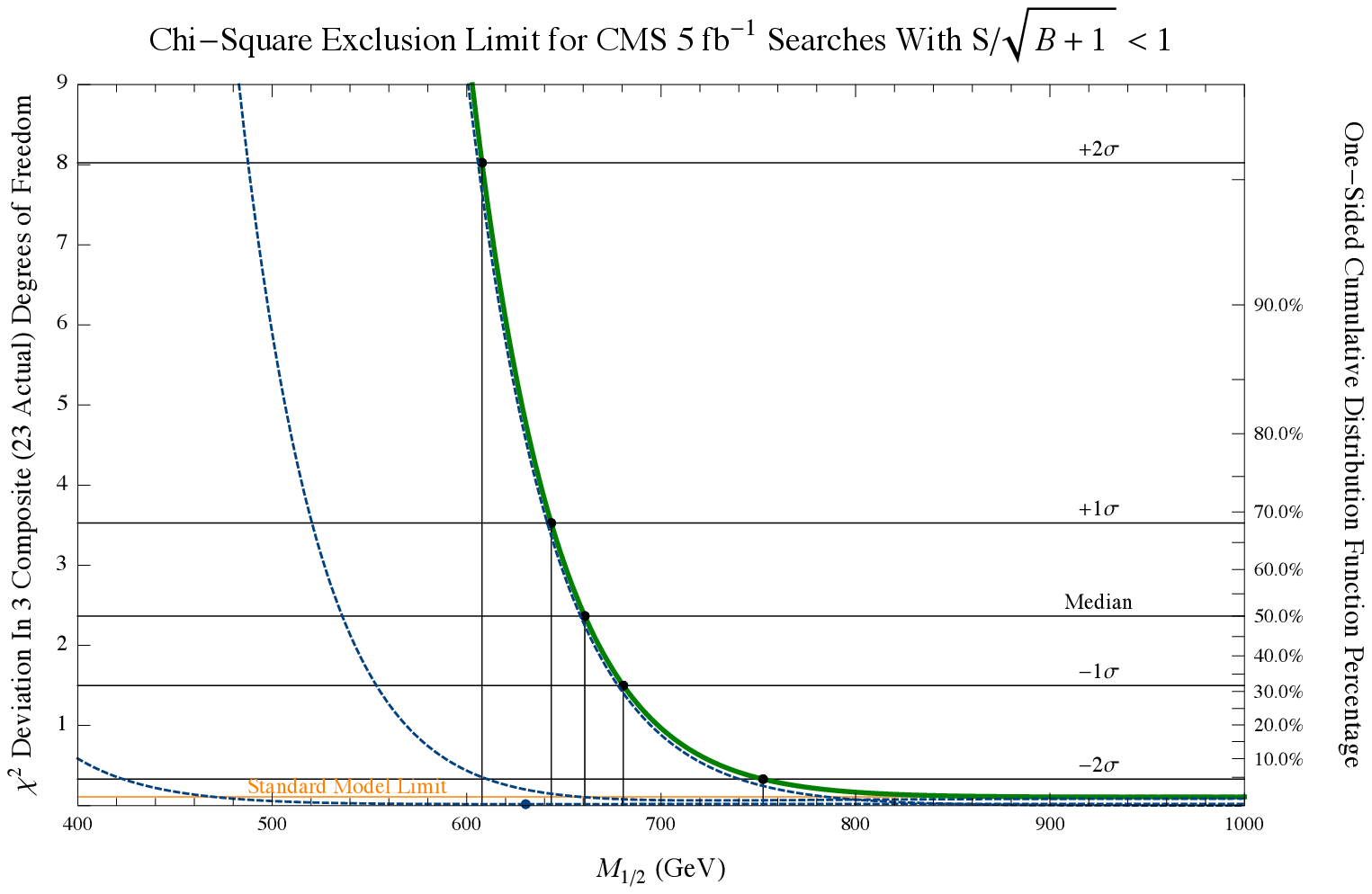}
        \caption{We depict a $\chi^2$ analysis of the 5 \fb class ATLAS (upper pane) and CMS (lower pane) studies
from Refs.~\cite{ATLAS-CONF-2012-033,ATLAS-CONF-2012-037,ATLAS-CONF-2012-041,CMS-PAS-SUS-11-020,Chatrchyan:2012qka,SUS-12-002}
that exhibit a signal significance $S/\sqrt{B+1}$ less than 1.  The thin dotted blue lines represent composite unit-strength
$\chi^2$ curves for each of the individual selections contained within the three search strategies analyzed for each collaboration, which are summed into the thick green cumulative multi-axis $\chi^2$ curve.
The intention of this study is the establishment of a lower bound on the \fsu5 SUSY mass scale.  At 2$\sigma$ (95\% confidence),
it appears that we may exclude gaugino masses $M_{1/2}$ below 608~GeV.  This is comfortably consistent with the best fit for
$M_{1/2}$ that is established by a parallel $\chi^2$ analysis of those searches exhibiting post-SM physics at a signal significance
greater than 1, as depicted in Figure (\ref{fig:CSBF}).}
        \label{fig:CSEXCL}
\end{figure*}

Our curiosity is piqued at the outset to note a correlation between the types of ATLAS and CMS searches that demonstrate event overproduction
in the first place.  In particular, we are interested in comparing the ATLAS hadronic observed statistics and signal significances of
Refs.~\cite{ATLAS-CONF-2012-037,ATLAS-CONF-2012-033} with those of the CMS hadronic $M_{T2}$ analysis of Ref.~\cite{SUS-12-002}.  Considering that both
ATLAS and CMS all-hadronic searches are targeting those multijet regimes that the \fsu5 model space is dominated by, the observed tracking in signal strength
is in keeping with what we must observe if the LHC is in fact producing \fsu5 supersymmetric events. Both the ATLAS hadronic search of Ref.~\cite{ATLAS-CONF-2012-033}
and the CMS $M_{T2}$ hadronic search of Ref.~\cite{SUS-12-002} are targeting squark and gluino pair-production through $\widetilde{q} \rightarrow q\widetilde{g}$ and $\widetilde{g} \rightarrow q \overline{q} \widetilde{\chi}_1^0$, albeit via independent discriminators, with ATLAS isolating the effective mass $M_{eff}$, while CMS employs the stransverse mass $M_{T2}$.
This compelling surface order continuity in the signal strength across these parallel all-hadronic studies has proved to be a precursor of the excellent
agreement observed in the deeper analysis, to which we next turn attention, of a comparative best fitting of the respectively favored SUSY particle mass scale.

We implement the multi-axis $\chi^2$ fitting procedure of Refs.~\cite{Li:2012hm,Li:2012tr} upon all the CMS 5 \fb analyses to date~\cite{CMS-PAS-SUS-11-020,Chatrchyan:2012qka,SUS-12-002}. For our 5 \fb $\chi^2$ fit, we partition all the CMS searches into two groups, one including only those 5 \fb searches that generate a minimum signal significance of \sb $>$ 1, and all the remaining CMS cases that cannot achieve this minimum threshold into a separate group. Likewise, we split all the ATLAS 5 \fb analyses completed thus far~\cite{ATLAS-CONF-2012-033,ATLAS-CONF-2012-037,ATLAS-CONF-2012-041} into one set with \sb $>$ 1, with the residual cases into a separate distinct set. We quadratically merge a statistical factor of $\sqrt{S+B+1}$ with the quoted collaboration estimates on the background uncertainty to account for Poisson fluctuations in the net experimental observation.  Searches demonstrating an anomalous under-production with respect to the data-driven background observations are zeroed out to allow the full error width for post-SM physics. All 5 \fb ATLAS and CMS SUSY searches are then evaluated against the entire \fsu5 model space presently consistent with all the latest experimental constraints, particularly the requirement of a 124-126 GeV Higgs boson mass, but excluding the ATLAS and CMS SUSY constraints that are the target of this analysis.
The narrow strip of otherwise viable model parameterizations ranging from 400 $\le M_{1/2} \le$ 900 GeV
is generously sampled at 22 representative benchmark combinations of $M_{1/2}$, $M_V$, $m_t$ and $\tan\beta$.

We execute on each of the 22 benchmark samples an in-depth Monte Carlo collider-detector simulation of all 2-body SUSY
processes based on the {\tt MadGraph}~\cite{Stelzer:1994ta,MGME} program suite, including the {\tt MadEvent}~\cite{Alwall:2007st},
{\tt PYTHIA}~\cite{Sjostrand:2006za} and {\tt PGS4}~\cite{PGS4} chain.
The SUSY particle masses are calculated with {\tt MicrOMEGAs 2.1}~\cite{Belanger:2008sj}, applying a proprietary modification of
the {\tt SuSpect 2.34}~\cite{Djouadi:2002ze} codebase to run the flippon-enhanced RGEs.
We implement a modified version of the default ATLAS and CMS detector specification cards provided with {\tt PGS4} that
calls a newly available anti-kt jet clustering algorithm, indicating an angular scale parameter
of $\Delta R = 0.4$ and $\Delta R = 0.5$, respectively.  The resultant event files are filtered according to a precise
replication of the selection cuts specified by the Collaborations, employing a script {\tt CutLHCO 2.0} of our own design~\cite{cutlhco}.
Lastly, the sampled event counts are utilized to extrapolate a continuous functional dependence on the gaugino mass $M_{1/2}$
that is suitable for the generation of a $\chi^2$ fitting of the \fsu5 event production against the experimental data. 

For the realization of b-tagging efficiencies in {\tt PGS4}, we maintain the default usage of fifth order polynomial fits,
though revising the numerical coefficients of the ``Loose'' b-tagging function as follows: $b\,(p_{\rm T}) = 0.0883 + 0.0197~p_{\rm T} - 2.4872 \times 10^{-4}~p_{\rm T}^2 +
1.47212 \times 10^{-6}~p_{\rm T}^3 - 4.16484 \times 10^{-9}~p_{\rm T}^4 + 4.41957 \times 10^{-12}~p_{\rm T}^5$ and
$b(\eta) = 1.00885 - 0.04975~\eta + 0.0693~\eta^2 - 0.03611~\eta^3 - 0.02222~\eta^4 + 0.00798~\eta^5$. The default ``Loose'' b-tagging functions in {\tt PGS4} have accordingly been shifted to the ``Tight'' b-tagging role. As a result, these new b-tagging functions process a ``Loose'' b-tag of about 60\% and a ``Tight'' b-tag of about 45\%. The ``Tight'' b-tag is applied to the replication of the CMS analysis of Ref.~\cite{SUS-12-002}, while the ``Loose'' b-tag is applied to the duplication of the CMS analysis of Ref.~\cite{CMS-PAS-SUS-11-020}.

In Ref.~\cite{Li:2012tr}, we engaged in a consistent cross-calibration of our Monte-Carlo quantitative procedure with that of the ATLAS Collaboration through normalization of a common mSUGRA benchmark. However, we remark that ATLAS and CMS exercise different schemes for derivation of the mSUGRA SUSY benchmark cross-section and related uncertainties. This is observed in practice in our Monte-Carlo results via no systematic suppression of event counts for CMS benchmark data, though in contrast a small systematic suppression is seen against ATLAS benchmark statistics, as accounted for in Ref.~\cite{Li:2012tr}.  This dichotomy in treatment of the common mSUGRA benchmarks between the two Collaborations obliges application of a uniform methodology to facilitate the fair comparison of results advertised by the two experiments against each other. As a step in this direction, we presently opt to institute a simple cross-calibration factor of 1.0 in the present work, permitting a genuine direct comparison of the ATLAS and CMS best fit SUSY mass scales for each \x2 fitting.  This decision is in keeping with our current focus
on the decidedly non-trivial relative correlation between the ATLAS and CMS SUSY search strategies with visible event excesses.
We note that this strategy represents a mild deviation from the ATLAS normalization factor estimate adopted in Ref.~\cite{Li:2012tr},
which we maintain to be an accurate reflection of the absolute SUSY spectrum coinciding with the cumulative $\chi^2$ minimum of the ATLAS search in isolation.
The \fsu5 best fit light stop mass $m_{\widetilde{t}_1}$ = 786 GeV, gluino mass $m_{\widetilde{g}}$ = 952 GeV and $u_R$ heavy squark mass $m_{\widetilde{u}_R}$ = 1371 GeV
established by that procedure are well corroborated by the ATLAS Ref.~\cite{ATLAS-CONF-2012-033} SUSY mass limits of $m_{\widetilde{g}}$ = 940 GeV and $m_{\widetilde{q}}$ = 1380 GeV.
As SUSY mass limits and anticipated SUSY spectrum measurements from both ATLAS and CMS are expected to converge,
we shall reanalyze the possible need for a non-unit mutual calibration factor in the future, as the delivered statistics continue to increase. 

We present in Figure (\ref{fig:CSBF}) the \x2 analyses focused on the No-Scale \fsu5 best fit of the mass parameter $M_{1/2}$ for the subset of overproductive ATLAS (upper pane) and CMS (lower pane) search strategies surmounting our minimum boundary for signal significance. The ATLAS upper pane of Figure (\ref{fig:CSBF}) illustrates the individual \x2 curves of 7j80 and 8j55 of Ref.~\cite{ATLAS-CONF-2012-037} and SRC Tight and SRE Loose of Ref.~\cite{ATLAS-CONF-2012-033}. The CMS lower pane of Figure (\ref{fig:CSBF}) depicts the \mt2 low $H_T$, \2b low $H_T$, and \mt2 high $H_T$ of Ref.~\cite{SUS-12-002}. This visually striking portrait in Figure (\ref{fig:CSBF}) exemplifies the intimate synchronization between the ATLAS and CMS hadronic searches when embedded within a No-Scale \fsu5 construction for our Universe. Apparent is the close proximity by which all searches that enjoy data excesses flourish, with all engendering a best fit $M_{1/2}$ in a narrow basin. Numerically speaking, we find a cumulative CMS \x2 best fit at about $M_{1/2}$ = 610 GeV, in concert with an ATLAS cumulative best fit at around $M_{1/2}$ = 655 GeV. A contiguous correlation such as this for parallel hadronic explorations is quite suggestive of the master framework that embodies the interconnection across experiments. The multi-axis cumulative \x2 best fits for $M_{1/2}$ between ATLAS and CMS exhibit a minor deviation of only about 7\%. This is a remarkable correlation when considering the fact that supersymmetry searches are from the squark and gluino production via the strong interaction. Furthermore, both cumulative \x2 wells in Figure (\ref{fig:CSBF}) also nicely correspond to the likewise uncalibrated \x2 minimum at $M_{1/2}$ = 610 GeV for the combination of 1--2 \fb ATLAS and CMS analyses in Ref.~\cite{Li:2012hm}. Therefore, not only do we see very non-trivial correlations across LHC experiments, but also across nearly five-fold increases in integrated luminosity. We believe it to be highly improbable that arbitrary fluctuations of the data-driven background are the source of the SUSY mass correlations coincidently affecting the same point in the targeted SUSY spectrum simultaneously in both LHC experiments.

Those searches deficient of sufficient excess events to satisfy our minimum threshold for signal significance report no information on viable production of a SUSY mass spectrum. Nevertheless, the amalgamation of all these null studies does establish a global lower bound on $M_{1/2}$ that requires consistency with SUSY production. So is the intent of Figure (\ref{fig:CSEXCL}), where 16 ATLAS (upper pane) and 23 CMS signal regions (lower pane) lacking compelling traces of new physics are included from the 5 \fb class reports in Refs.~\cite{ATLAS-CONF-2012-033,ATLAS-CONF-2012-037,ATLAS-CONF-2012-041,CMS-PAS-SUS-11-020,Chatrchyan:2012qka,SUS-12-002}, all registering a signal significance \sb $<$ 1. In Figure (\ref{fig:CSEXCL}), we adopt a single-sided cumulative distribution function, and are interested in the values of $M_{1/2}$ at the median, $+1 \sigma$ and $+2\sigma$ intersections of the $\chi^2$ statistic. Anticipating strong interdependence amongst parallel event selections within a single SUSY search strategy, we condense each family of selections under investigation into a single unit-strength composite degree of freedom. Evident in Figure (\ref{fig:CSEXCL}) is that we may exclude values of $M_{1/2} <$ 614 GeV for ATLAS and $M_{1/2} <$ 608 GeV for CMS at 2$\sigma$ (95\% confidence) level. The 1$\sigma$ and median intersections with the $\chi^2$ curve for ATLAS occur at 653 and 671 GeV, respectively, and at 644 and 661 GeV for CMS. These ranges offer a very suitable overlap with the intersection boundaries of the median fit for the \x2 wells of Figure (\ref{fig:CSBF}), which occur at 603 and 769 GeV for ATLAS, and at 560 and 754 GeV for CMS.

The patent correlations introduced here amid both ATLAS and CMS experiments set up the possibility of a compelling confirmation in 2012 when the $\sqrt{s}$ = 8 TeV collision data begins arriving. With an expected yield of 15 \fb in 2012 at this increased beam energy, the question of SUSY production or random background fluctuation will surely be resoundingly answered. With signal significances of the ATLAS and CMS searches explored in this work foreseen to approach and surpass the gold standard of \sb $>$ 5 in 2012, if in fact we inhabit a No-Scale \fsu5 universe, there will be little doubt as to the fate of the alluring scent of a signal that is already effusing from the first year of LHC collider operation data.

\textbf{Conclusions}--In the time since proton-proton collisions initially began at the LHC in 2010, the mantra from the ATLAS and CMS experiments has been that there
remains no evidence for supersymmetry.  With early elevated expectations for a rapid and decisive discovery dashed, the demeanor of SUSY enthusiasts has become ensnared of late
in a downward spiral tracking the lack of proof for SUSY at the LHC. This may be about to change.

We presented persuasive signs that confirmation of SUSY may in fact be just around the proverbial corner, thanks to an increased beam energy of 8 TeV. The obstruction all along may not have been an absence of SUSY in nature, but the exclusion of the naturally ubiquitous framework for our Universe from the experimental surveys of the supersymmetric landscape. Once the parallel hadronic searches of ATLAS and CMS are immersed within a No-Scale \fsu5 structure, the SUSY mass spectrum and beautiful correlations across the LHC experiments appear to come to life.

Supplementing our previous work on ATLAS all-hadronic and multijet SUSY searches with newly available statistics from CMS also employing all-hadronic cuts in conjunction with the stransverse mass variable \mt2, we uncovered through a fresh \x2 fitting procedure that corresponding 5 \fb ATLAS and CMS searches are registering excess events that indeed do correlate to both experiments in the range of $M_{1/2}$ = 610--655 GeV. This SUSY mass scale further precisely matches our prior combined ATLAS and CMS 1--2 \fb best fit SUSY mass of $M_{1/2}$ = 610 GeV. The congruence of these findings across both LHC experiments renders as increasingly improbable an attribution of the observed excesses to coincidental random fluctuations of the data-driven background at the same SUSY mass scale.

We and many within the SUSY exploration community fervently await the next tranche of 8 TeV LHC collision data, presumed to be a minimum of an additional 5 ${\rm fb^{-1}}$, and possibly more. If the early distant warning reverberating from the non-trivial correlations presented here is any indication, this next synthesis of ATLAS and CMS observations could resonate even louder.

%%%%%%%%%%%%%%%%%%%%%%%%%%%%%%%%%%%%%%%%%%%%%%%%%%%%%%%%%%%%%%%%%%%%%%%%%%%%

\begin{acknowledgments}
This research was supported in part
by the DOE grant DE-FG03-95-Er-40917 (TL and DVN),
by the Natural Science Foundation of China
under grant numbers 10821504, 11075194, and 11135003 (TL),
by the Mitchell-Heep Chair in High Energy Physics (JAM),
and by the Sam Houston State University
2011 Enhancement Research Grant program (JWW).
We also thank Sam Houston State University
for providing high performance computing resources.
\end{acknowledgments}

%%%%%%%%%%%%%%%%%%%%%%%%%%%%%%%%%%%%%%%%%%%%%%%%%%%%%%%%%%%%%%%%%%%%%%%%%%%%

\bibliography{bibliography}

\end{document}